
%
%
%
\let\SELECTOR=P      
%
%
%
\expandafter\ifx\csname phyzzx\endcsname\relax \input phyzzx \fi
%
%
\interdisplaylinepenalty=10000

%
%
\newdimen\doublewidth
\doublewidth=12in
\newinsert\LeftPage
\count\LeftPage=0
\dimen\LeftPage=\maxdimen
\def\PageBox{\vbox{\makeheadline \pagebody \makefootline }}
\def\papersize{\hsize=412pt \vsize=570pt \pagebottomfiller=0pt
    \skip\footins=\bigskipamount \normalspace }
\papersize
\if R\SELECTOR
    \mag=833
    \voffset=-0.3truept
    \hoffset=-0.5truein
    \output={\ifvoid\LeftPage \insert\LeftPage{\floatingpenalty 20000 \PageBox}
        \else \shipout\hbox to\doublewidth{%
            \box\LeftPage \hfil \PageBox }\fi
        \advancepageno
        \ifnum\outputpenalty>-20000 \else \dosupereject \fi }
    \message{Warning: some DVI drivers cannot handle reduced output!!!}
\else
    \mag=1000
    \voffset=\VOFFSET
    \hoffset=0pt
\fi
%
%

%
%
\newif\iffigureexists
\newif\ifepsfloaded
\openin 1 epsf
\ifeof 1 \epsfloadedfalse \else \epsfloadedtrue \fi
\closein 1
\ifepsfloaded \input epsf \fi
\def\checkex#1 {\relax \ifepsfloaded \openin 1 #1
        \ifeof 1 \figureexistsfalse \else \figureexiststrue \fi
        \closein 1
    \else \figureexistsfalse \fi }
\def\diagram#1 {\vcenter{\def\epsfsize##1##2{0.4##1} \epsfbox{#1}}}
%
%
\def\del{\partial}

\def\smallmat#1#2#3{\left( \vcenter{\baselineskip=12pt
        \ialign{\hfil$\scriptstyle{##}$&\kern 8pt\hfil$\scriptstyle{##}$\cr
                #1& #2\cr #2& #3\cr }}\right)}

\def\nv{n_V}
\def\nh{n_H}
\def\h{h}
\def\FNP{h^{\rm np}}
\def\Fh{F^{\rm het}}
\def\Ftwo{F^{\rm II}}

\def\A{A}
\def\Sinv{S^{\rm inv}}
\def\g{g}
\def\gr{g_{\rm grav}}
\def\br{b_{\rm grav}}
\def\LL{L}
\def\mpl{M_{\rm Pl}}
%
%

%
%

\def\sltwo{SL(2,{\bf Z})}

%
%
\def\ldf{\REF}
\def\JJjournal#1#2{\unskip\space{\sfcode`\.=1000 #1 \bf #2}\space }
\def\nup#1 {\JJjournal{Nucl. Phys.}{B#1}}
\def\plt#1 {\JJjournal{Phys. Lett.}{#1}}
\def\cmp#1 {\JJjournal{Comm. Math. Phys.}{#1}}
\def\prp#1 {\JJjournal{Phys. Rep.}{#1}}
\def\prl#1 {\JJjournal{Phys. Rev. Lett.}{#1}}
\def\prev#1 {\JJjournal{Phys. Rev.}{#1}}
\def\mplt#1 {\JJjournal{Mod. Phys. Lett.}{#1}}
%
%
%
\ldf\SW{N.~Seiberg and E.~Witten, Nucl. Phys. {\bf B426} (1994) 19,
  Nucl. Phys. {\bf B431} (1994) 484.}
\ldf\Ceresole{A. Ceresole, R. D'Auria and  S. Ferrara,
Phys.~Lett.~{\bf B339} (1994) 71;
A. Ceresole, R. D'Auria, S. Ferrara and A.
Van Proeyen, preprints hep-th/941220, hep-th/9502072.}
\ldf\KLTY{A.~Klemm, W.~Lerche, S.~Theisen and  S.~Yankielowicz,
  Phys.~Lett.~{\bf B344} (1995) 169.}
\ldf\KV{S.~Kachru and C.~Vafa, preprint hep-th/9505105.}
\ldf\FHSV{S. Ferrara, J.~Harvey, A.~Strominger and C.~Vafa,
preprint hep-th/9505162.}
\ldf\seiberg{N.~Seiberg, Nucl. Phys. {\bf B303} (1988) 286;
S. Cecotti, S. Ferrara and L. Girardello,
Int. J. Mod. Phys. {\bf A4} (1989) 2457;~
D. L\"ust and S. Theisen, Int. J. Mod. Phys. {\bf A4} (1989) 4513.}
\ldf\ntwo{B.\ de Wit and A.\ Van Proeyen, \nup245 (1984) 89;~
B.\ de Wit, P.\ Lauwers and A.\ Van Proeyen, \nup255 (1985) 569;~
E.\ Cremmer, C.\ Kounnas, A. \ Van Proeyen, J.P.\ Derendinger,
S.\ Ferrara, B.\ de Wit and L.\ Girardello,  \nup250 (1985) 385.}
\ldf\CDFKM{ P.~Candelas, X.C.~de la Ossa, A.~Font, S.~Katz and D.~Morrison,
\nup416 (1994) 481.}
\ldf\HKTY{S.~Hosono, A.~Klemm, S.~Theisen and S.T.~Yau,
                \cmp 167 (1995) 301.}
\ldf\DKLL{B.\ de Wit, V.~Kaplunovsky, J.~Louis and D.~L\"ust,
preprint hep-th/9504006.}
\ldf\AFGNT{I.~Antoniadis, S.~Ferrara, E.~Gava, K.~Narain and T.~Taylor,
     preprint hep-th/9504034.}
\ldf\NRT{M.~Grisaru and W.~Siegel,  Nucl. Phys. {\bf B201} (1982) 192;~
  P.~Howe, K.~Stelle and P.~West, Phys.~Lett.~{\bf 124B} (1983) 55;~
 N.~Seiberg,  Phys.~Lett.~{\bf B206} (1988) 75.}
\ldf\CLM{G.L.~Cardoso, D.~L\"ust and T.~Mohaupt, preprint hep-th/9412209.}
\ldf\HKTYII{S. Hosono, A. Klemm, S. Theisen and S.T. Yau, \nup433 (1995) 501.}
\ldf\CDGP{P.~ Candelas, X.C.~de la Ossa, P.S.~Green and
            L.~Parkes, \plt B258 (1991) 118, \nup359 (1991) 21.}
\ldf\Oleg{O. Ogievetsky, unpublished.}
\ldf\weyl{M.~de Roo, J.W.~Van Holten, B.\ de Wit and A.\ Van Proeyen,
\nup173 (1980) 175;~
E.~Bergshoff, M.~de Roo and B.\ de Wit, \nup182 (1981) 173.}
\ldf\BCOV{M.~Bershadsky, S.~Cecotti, H.~Ooguri and C.~Vafa, \nup 405 (1993)
279;
         \cmp 165 (1994)  311.}
\ldf\anom{L.~Dixon, V.~Kaplunovsky and J.~Louis, \nup355 (1991) 649;~
       J. Louis, in the Proceedings of the
        {\it 2nd International Symposium on Particles, Strings and Cosmology},
        Boston, MA, March 25-30, 1991, ed.~P.~Nath und S.~Reucroft;~
G.L.~Cardoso  and B. Ovrut, \nup369 (1992) 351; \nup392 (1993) 315;~
J.P.~Derendinger, S.~Ferrara, C.~Kounnas and F.~Zwirner, \nup372 (1992) 145;~
L.~Ib\'a\~nez and D.~L\"ust, \nup382 (1992) 305;~
V.~Kaplunovsky and J.~Louis, \nup422 (1994) 57; preprint hep-th/9502077.}
\ldf\AGN{I.~Antoniadis, E.~Gava and K.~Narain,
\plt B283 (1992) 209; \nup383 (1992) 93;
I.~Antoniadis, E.~Gava,  K.~Narain and T.~Taylor, \nup413 (1994) 162.}


\Pubnum={LMU-TPW 95--9\cr
         UTTG--12--95}
\date={June 1995}
\titlepage
\title{Aspects of Duality in $N=2$ String Vacua
        \foot{Research supported in part by:
        the NSF, under grant PHY--90--09850 (V.K.);
        the Robert A.~Welch Foundation (V.K.);
        the Heisenberg Fellowship of the DFG (J.L.);
        the NATO, under grant CRG~931380 (V.K.\&J.L.);
        the German-Israeli
        Foundation for Scientific Research  (J.L.\&S.T.).}}
\author{Vadim Kaplunovsky,
        \foot{Email: \tt vadim@bolvan.ph.utexas.edu}}
\address{Theory Group, Dept.~of Physics, University of Texas\break
        Austin, TX 78712, USA}
\author{Jan Louis
        \foot{Email: \tt jlouis@lswes8.ls-wess.physik.uni-muenchen.de}
 \quad and \quad  Stefan Theisen
  \foot{Email: \tt theisen@mppmu.mpg.de}
}
\address{Sektion Physik, Universit\"at M\"unchen\break
        Theresienstrasse 37, D-80333 M\"unchen, Germany}
\abstract
We collect further evidence for the proposed duality between $N=2$ heterotic
and type II
string vacua in a specific model suggested by Kachru and Vafa.
In the gauge sector the previous analysis is extended;
it is further shown that the duality also holds for the one--loop gravitational
couplings to the vector multiplets.
\endpage
%
%
1.  The recent advances in understanding non-perturbative
aspects of $N=2$ supersymmetric Yang-Mills theories [\SW] have raised the
question whether similar techniques are applicable in string theory.
One of the key elements in the work of Seiberg and Witten is the
relation between a gauge theory at strong coupling
and a `dual' theory at weak coupling with magnetic monopoles (dyons)
as elementary excitations. This dual theory  can be analyzed in perturbation
theory and, as a consequence,
the exact non-perturbative low energy effective action for all values
of the coupling constant is determined.

In order to apply such techniques to string theory, a similar strong -- weak
coupling duality has to be established.  For $N=2$ string theories
it has been conjectured that such a duality exists between
heterotic vacua compactified on the  six-dimensional manifold
$K_3 \times T^2$ and
type II vacua compactified on a Calabi-Yau threefold [\Ceresole--\FHSV].
The coupling constant in string theory is a dynamical variable
determined by the
vacuum expectation value of the dilaton $S$. In $N=2$ heterotic vacua
$S$ resides  in an abelian vector multiplet
while in type II vacua it is a member of  a hypermultiplet [\seiberg].
Combining the facts that there are no gauge neutral couplings between
vector and hypermultiplets [\ntwo] and that $S$ organizes the string
perturbation theory implies
a non-renormalization theorem for both,  heterotic and type II vacua.
For the heterotic  vacua the tree level couplings of the
hypermultiplets are exact
whereas the tree level couplings of the vector multiplets
are corrected at one-loop and non-perturbatively.
In type II vacua the situation is reversed and the tree level couplings of the
vector multiplets are exact while the hypermultiplets suffer perturbative
and non-perturbative corrections.
Thus, if a string vacuum has a dual representation as both heterotic and
type II the exact effective Lagrangian can  be obtained by computing
the couplings of the vector multiplets in the type II theory
and the couplings of the hypermultiplets in the heterotic theory.

Concrete examples of  `dual pairs' of $N=2$ string vacua
have  been suggested in refs.~[\KV,\FHSV] and non-trivial
evidence for the proposed duality was found.
One of the models considered in ref.~[\KV] is a specific compactification
of the heterotic string on  $K_3\times T^2$ with gauge group
$U(1)^3$.
The corresponding gauge bosons are the graviphoton, the vector partner of the
dilaton and the vector partner of the toroidal modulus $T$. The
second toroidal modulus $U$ is locked at $U=T$.
Thus the model has two $U(1)$ vector multiplets $(\nv=2)$ while
the number of hypermultiplets turns out to be $\nh = 129$.
Kachru and Vafa observe that there is an unique
Calabi--Yau threefold  $X_{12}(1,1,2,2,6)$, ---  the degree 12 hypersurface
in ${\bf P}^4(1,1,2,2,6)$ ---
with $\nv = b_{(1,1)} = 2$ and $\nh = b_{(1,2)} +1  = 129$
($b_{(1,1)} $ and $b_{(1,2)}$ denote the number of $(1,1)$ and $(1,2)$ forms
and the `+1' counts the dilaton).
Therefore this Calabi-Yau space is a good candidate for the dual
type IIA string vacuum.
The tree level couplings of the two vector multiplets
are  known exactly for the type II vacuum [\CDFKM,\HKTY] while
for the dual heterotic vacuum
they have only been studied in perturbation theory [\Ceresole,\DKLL,\AFGNT].
Kachru and Vafa have given evidence that they agree at weak coupling.
In this letter we extend
their analysis of the gauge sector and further show that also for
the gravitational coupling to vector multiplets the duality between
the vacua holds.
Our  results in section 2 overlap with recent work of K. Narain and
collaborators.

2.  The couplings of the vector multiplets are encoded
in a holomorphic prepotential $F$ [\ntwo].
In the heterotic vacuum $F^{\rm het}$ has the weak coupling expansion
$$
\Fh = \half S T^2 + \h(T) + \FNP (e^{-8\pi^2 S},T)\ ,
\eqn\Fhet
$$
where
$\half S T^2$ is the tree level contribution,
$\h(T)$ is the dilaton independent one-loop correction and
$\FNP$ is generated non-perturbatively.
  \foot{The standard $N=2$ non-renormalization theorem states that beyond
          one-loop there are no further perturbative corrections [\NRT].}
In ref.~[\DKLL,\AFGNT] it was shown that $\h(T)$
is strongly constrained by its transformation properties under any
exact quantum symmetry and by its singular behaviour at special points in the
moduli space where additional massless states appear.
The model at hand has an exact $\sltwo$ quantum symmetry
($T$-duality) which acts on the modulus $T$ according to
$$
T \rightarrow {a T - ib \over icT +d}\,,
\qquad \pmatrix{a&b\cr c&d\cr}\,\in\,\sltwo,
\eqn\modtrans
$$
while the dilaton $S$ is invariant at the tree level.
Using the formalism developed in ref.~[\DKLL] it is
straightforward to determine the transformations law of $\h$
under this $\sltwo$
$$
\h(T) \rightarrow {\h(T) \over (icT +d)^4}\  +\  {\Xi (T)  \over (icT +d)^4}\ ,
\eqn\htrans
$$
where  $\Xi $ is at most a quartic polynomial in $T$
arising from the multivaluedness
of $\h(T)$. (In the absence of logarithmic singularities
$\h$ would be a modular form of weight $-4$.)
The  5th derivative $\del_T^5\h(T)$
does not suffer from any ambiguity
and is a modular form of weight +6.
  \foot{Note that the n-th ordinary derivative $\del_T^n f_{1-n}$
       of  a weight $n-1$ modular  form $f_{1-n}$
       is  a  modular  form of weight $n+1$.}
The singularities of $\h$ are at $T=1,\infty$;
at $T=1$ the gauge group $U(1)^3$ is enlarged to $SU(2)\times U(1)^2$
(with no additional massless hypermultiplets)
and, as a consequence, $\del_T^2 \h$ develops a logarithmic singularity
$\del_T^2 \h \sim - {b_{SU(2)}\over 8\pi^2} \ln (T-1)$.
 \foot{A more extensive discussion of the singularities of $\h$
 and its precise relation to the gauge coupling is given in
 refs.~[\CLM,\DKLL,\AFGNT].}
Hence, modular invariance (together with $b_{SU(2)}= -4$) dictates
$$
\del_T^2 \h = {1\over 4\pi^2} \ln [j(iT)-j(i)] + {\rm finite\ terms} \ ,
\eqn\hj
$$
where $j(iT)$ is the modular invariant $j$-function.

At $T=\infty$ the coupling $\del_T^2\h$ should diverge at most
like $T^2$ which implies that $\del_T^5 \h(T)$ is regular
everywhere except at $T\sim1$ where
$
\del_T^5 \h \sim  \pi^{-2}  (T-1)^{-3} .
$
This determines
$\del_T^5 \h(T)$ up to one arbitrary coefficient \A
$$
\del_T^5 \h\ =\ \pi\left[{E_4^6\over E_6^3 }\, +\,\A\,{E_4^3\over E_6 }\,
-\,(\A+1)\,E_6\right] \ ,
\eqn\hresult
$$
where $E_4, E_6$ are the normalized Eisenstein functions of weight $4,6$,
respectively. To determine the coefficient \A\ we use the results of
toroidal compactifications where, in addition to $T$, the modulus
 $U$ is also unconstrained. For this case the
third derivatives $\del_T^3 \h(T,U),\,\del_U^3\h(T,U)$ have been determined in
refs.~[\DKLL,\AFGNT]. In terms of the coordinates $\phi^\pm \equiv T\pm U$,
it is possible to compute $\del_{\phi^{+}}^5 \h(\phi^+,\phi^-)$ at $\phi^-=0$
from the knowledge of the third derivatives
$\del_T^3 \h(T,U),\,\del_U^3\h(T,U)$.
$\del_{\phi^{+}}^5 \h(\phi^+,0)$  is singular at $T=1$
(where the enhanced gauge symmetry is $SU(2)^2$)
and at $T=e^{i \pi/6}$ (where the enhanced gauge symmetry is $SU(3)$);
nevertheless the singularity at $T=1$ has to agree with eq.~\hresult\
since at that point the additional massless states which contribute to the
$\beta$-function are identical in both theories.
Matching the coefficients of the singular terms at $T=1$ yields $A=-{23/18}$.

The analysis of refs.~[\DKLL,\AFGNT]  also showed that at  one-loop the
dilaton $S$ is no longer invariant under $T$-duality.
Instead it transforms according to
$$
S \rightarrow S\  -\ {1\over 3} \del^2_T \Xi\
+\ 2 i c\, {\del_T (\h + \Xi) \over (icT+d)}\
      +\ 4 c^2\, { \h +  \Xi \over (icT+d)^2}\
      +\ i\, {\rm const.} \,.
\eqn\Strans
$$
It is however possible to define a modular invariant coordinate  by
$$
\Sinv := S + \coeff13 \left[\del_T^2 \h(T) + \LL (T) \right] \ ,
\eqn\Sinvdef
$$
where the holomorphic $\LL(T)$ is modular invariant
up to a shift by an imaginary constant.
The difference $\Sinv - S$ has to be finite f
or finite $T$ and should not grow faster
than a polynomial at $T\rightarrow \infty$.
Therefore, eq.~\hj\ determines
$$
\LL =   - {1\over 4\pi^2} \ln [j(iT)-j(i)]  + {\rm const.}\ .
\eqn\Ldef
$$
It is important to note that $S$ is a $N=2$ special coordinate
but $\Sinv$ is {\it not}.

Let us now turn to the type II string compactified on the
Calabi--Yau threefold
$X_{12}(1,1,2,2,6)$.
The defining polynomial of the mirror manifold (which has $b_{(1,2)}=2$)
is given by  [\CDFKM,\HKTY]
$$
p = z_1^{12} +   z_2^{12} +   z_3^6 +   z_4^6 +   z_5^2
+ a_0 z_1 z_2 z_3 z_4 z_5 +a_1  z_1^6  z_2^6 \ ,
\eqn\pdef
$$
where $a_0$ and $a_1$ are the two complex structure deformations.
The uniformizing variables at large complex structure are
$x={a_1\over a_0^6}$ and $y={1\over a_1^2}$ and
the manifold \pdef\ has a conifold singularity  at
$$
(1-12^3 x)
^2 -4\cdot 12^3\, x^2 y = 0\ .
\eqn\conifold
$$
For generic values of $y$ this is satisfied by two values of $x$, which
coalesce for $y=0$. This observation, which is reminiscent of the work of
Seiberg and Witten [\SW], led Kachru and Vafa to identify $y=0$ with the
weak coupling limit of the dual heterotic vacuum.

In order to make this proposal more precise one has to find
a map between the special coordinates $S$ and $T$ in the heterotic
vacuum and  the $(1,1)$ deformations of the Calabi-Yau manifold.
The special coordinates $t_1$ and $t_2$ on $X_{12}(1,1,2,2,6)$
are determined by the mirror map
in terms of $x$ and $y$ [\CDFKM,\HKTY].
The mirror map can be inverted, leading at weak coupling to
$$
x = {1\over j(q_1)}  + O(q_2) ,\qquad
y = \,  q_2\   \g(q_1) + O(q_2^2) \, ,\qquad
q_j=e^{2\pi i t_j}\,,
\eqn\xyweak
$$
where $g(q_1)$ is a power series in $q_1$, normalized to $g(0)=1$.
The first few coefficients are recorded in [\CDFKM] or can be computed
using the computer program of [\HKTYII].
One is now led to the following identification of the special
coordinates in the two vacua:
$t_1 = iT,\,  t_2 =  4\pi i S$.
 \foot{Note that eq.~\Strans\ implies that $S$ is ambiguous
       up to a quadratic polynomial in $T$. We acknowledge
       usefull discussions with P. Mayr on that point.}

Once the coordinates have been identified one has to check the identity
of the two prepotentials $\Fh=\Ftwo$.
For the Calabi-Yau manifold the Yukawa couplings $Y_{ijk}$ are computed
 in refs.~[\CDFKM,\HKTY] and when expressed in terms of special coordinates
they  are determined by the third derivative of the prepotential
$Y_{ijk}= \del_{t_i}  \del_{t_j} \del_{t_k} \Ftwo$.
Using the formuli of [\CDFKM,\HKTY] it is straightforward to compute
in the weak coupling limit ($y\to 0$)
$$
\del^3_T  \Ftwo ={1\over4\pi^2}{E_4 \over \omega_0^2}\  \del_T
\left( \ln[j(iT) - j(i)] - \coeff32 \ln\g (iT) \right) \,  ,\qquad
\del^2_T \del_S  \Ftwo = {E_4 \over \omega_0^2} \,  ,
\eqn\yukawa
$$
while $\del^2_S \del_T  \Ftwo$ and $ \del^3_S  \Ftwo$ vanish in this limit.
Here we have chosen a convenient overall normalization of $\Ftwo$.
$\omega_0$ is the fundamental period of the Calabi-Yau manifold
and it appears in the transformation of the Yukawa couplings
to special coordinates  [\CDGP];
it plays the role of the homogeneous $N=2$ coordinate $X_0$.
The two equations in \yukawa\  are  consistent if
$$
\omega_0^2 (x, y=0)\
\equiv\left(\sum_{n=0}^\infty{(6n)!\over(3n)!(n!)^3}\,j(iT)^{-n}\right)^2
=\, E_4(iT)\,,
\eqn\omegae
$$
where the first equation follows from the explicit series
representation of $\omega_0$ [\CDFKM,\HKTY].
We have checked this
identity perturbatively in $q_1$; an analytic proof has been given by
O. Ogievetsky [\Oleg].
Inserting eq.~\omegae\ into \yukawa\ we  arrive at
$$
\del^2_T  \Ftwo =
S  + {1\over 4\pi^2} \ln[j(iT) - j(i)] - {3\over 8\pi^2}  \ln\g (iT) \  .
\eqn\ttresult
$$

Now we are prepared to compare the two prepotentials.
Equating  \Fhet\ with \ttresult\
($\Fh=\Ftwo$)   implies
$$
\del^2_T  \h = {1\over 4\pi^2} \ln[j(iT) - j(i)]
- {3\over 8\pi^2} \ln\g (iT)\  ,
\eqn\tthrelation
$$
where $\g(iT) $ has been defined  in eq.~\xyweak.
Inserting \tthrelation\ into \hresult\ now facilitates a non-trivial
check of the consistency of the proposed duality.
We were able to verify the
consistency  of eqs.~\tthrelation\ and  \hresult\ up to order 10 in $q_1$.
Furthermore, inserting eqs.~\tthrelation, \Ldef\ into \Sinvdef\
we find
$$
\Sinv = S - {1\over 8\pi^2}  \ln\g (iT) \ , \qquad \Rightarrow  \qquad
e^{-8\pi^2 \Sinv} =  e^{-8\pi^2 S} \g(iT) \equiv y \  .
\eqn\Sinvdef
$$
Hence,  at leading order the Calabi-Yau coordinate $y$ precisely
corresponds to the invariant dilaton defined by eq.~\Sinvdef\
and therefore is modular invariant.

3.  So far we have concentrated on the duality in the gauge couplings
of the two vacua.
It is possible to  extend the analysis
and  show the duality also between the
gravitational couplings of the  vector multiplets.
In $N=2$ supergravity a particular combination of
higher derivative curvature terms (including $R\tilde R$)
reside in the square of the (chiral) Weyl  superfield [\weyl].
Its coupling to the (abelian) vector multiplets is governed by
a holomorphic  function $F_1$.
In type II vacua $\Ftwo_1$ is only generated at the one-string loop
level  and in ref.~[\BCOV] a general prescription for its
computation in terms of topological amplitudes was given.
For the particular Calabi--Yau threefold $X_{12}(1,1,2,2,6)$
$\Ftwo_1$ has the expansion
$$
\Ftwo_1 = - {2\pi i \over 12} (52 t_1 + 24 t_2)
+ \sum_{jk} \left[2d_{jk} \ln  \eta_0(q_1^j q_2^k) +
\coeff16 n_{jk} \ln(1- q_1^j q_2^k)\right]\ ,
\eqn\Fonettwo
$$
 where $\eta_0(q)= \prod_1^\infty (1-q^n)=q^{-1/24}\eta(q)$.
The first few coefficients $d_{jk},n_{jk}$ have been
 explicitly computed  in ref.~[\CDFKM] .

If the proposed duality is to hold it should be possible to identify
the same coupling also in the heterotic vacuum.
In complete  analogy with eq.~\Fhet\
 $\Fh _1$  has a weak coupling expansion
$$
\Fh_1 = 24\,  S + \h_1 (T) \ +\  {\rm non-perturbative} \ ,
\eqn\Fonepert
$$
where the factor of 24 is the standard normalization of the
curvature couplings.
As before  $\h_1 (T)$ is strongly constrained by its
modular properties and its singularities on the moduli space.
Near $T\sim 1$ the singular contribution  to $\h_1 (T)$
coincides with the correction  for $\del^2_T\h$
since no additional gauge singlets become massless
and we have
 $$
 \h_1 =  {1\over 4\pi^2} \ln [j(iT)-j(i)] + {\rm finite\ terms} \ .
\eqn\honej
$$
On the other hand the modular transformation properties of  $\h_1 (T)$
are determined from the holomorphic or modular anomaly of this coupling
[\anom,\AGN,\BCOV].
Repeating the analysis for the gravitational couplings of the vector
multiplets we find that the  non-holomorphic coupling
$$
\gr^{-2}
=\ Re  \Fh_1 (S,T)\
+\ {\br\over 16\pi^2}\,
\Big(\log{\mpl^2\over p^2} +  K(S,T)\Big)
\eqn\anofin
$$
has to be modular invariant.
Here $\br = 2 [\nh - (\nv-1) +22]=300$ is the one-loop
coefficient of the `gravitational $\beta$-function' [\AGN]  and $K$
is the K\"ahler potential given by
$$
\eqalign{
K=\ & - \ln(S+\bar S - V_{\rm GS}) - 2\, \ln (T+\bar T)\ , \cr
{\rm where}\quad &
V_{\rm GS} = 4\, (T + \bar T)^{-2} (\h +\bar{\h})
- 2\, (T + \bar T)^{-1} (\del_T \h + \del_{\bar T} \bar{\h}) \ .}
\eqn\Kpot
$$
The requirement of keeping $\gr^{-2}$ modular invariant
uniquely determines
$$
\Fh_1 = 24\,  \Sinv + {1\over 4\pi^2} \ln [j(iT)-j(i)]
- {300\over 4\pi^2} \ln \eta^2(iT)\ .
\eqn\Fonehet
$$
We compared eqs.~\Fonehet\ and \Fonettwo\
(using an appropriate normalization) as a power series in $q_1$
and found agreement
up to fourth order.
This is a further check of the duality between the two
string vacua.

\refout
\end